\documentclass[11pt,nofootinbib]{revtex4-1}
\usepackage{amssymb}

\begin{document}

\title{Study of stability of relativistic ideal Bose-Einstein condensates}

\author{F. Briscese$^{ab}$}\email{fabio.briscese@sbai.uniroma1.it}
\author{M. Grether$^{c}$} \author{M. de Llano$^{d}$}
\author{George A.Baker, Jr.$^{e}$}

\affiliation{$^{a}$ Istituto Nazionale di Alta Matematica
Francesco Severi, Gruppo Nazionale di Fisica Matematica,
Citt$\grave{a}$ Universitaria, P.le A. Moro 5, 00185 Rome,
Italy.\\
$^{b}$ DSBAI, Sezione di Matematica, Sapienza Universit$\grave{a}$
di Roma, Via Antonio Scarpa 16,  00161 Rome,  Italy.\\
$^{c}$Facultad de Ciencias, Universidad Nacional Aut\'{o}noma de
M\'{e}xico, 04510 M\'{e}xico, DF, MEXICO\\
$^{d}$Instituto de Investigaciones en Materiales, Universidad
Nacional Aut\'{o}noma de M\'{e}xico, 04510 M\'{e}xico, DF, MEXICO\\
$^{e}$Theoretical Division, Los Alamos National Laboratory, Los
Alamos, NM 87545, USA}

\begin{abstract}
A relativistic complex scalar boson field at finite temperature $T$ is
examined below its critical Bose-Einstein condensation temperature. It is
shown that at the same $T$\ the state with antibosons has higher entropy,
lower Helmholtz free energy and higher pressure than the state without
antibosons---but the same Gibbs free energy as it should. This implies that
the configuration without antibosons is metastable. Results are generalized
for arbitrary $d$ spatial dimensions.

PACS \# 67.85.Hj; 67.85.Jk; 03.75.Kk; 05.30.Jp; 47.75.+f
\end{abstract}

\maketitle

\affiliation{$^{a}$ Istituto Nazionale di Alta Matematica
Francesco Severi, Gruppo Nazionale di Fisica Matematica,
Citt$\grave{a}$ Universitaria, P.le A. Moro 5, 00185 Rome,
ITALY\\
$^{b}$ DSBAI, Sezione di Matematica, Sapienza Universit$\grave{a}$
di Roma, Via Antonio Scarpa 16,  00161 Rome,  ITALY\\
$^{c}$Facultad de Ciencias, Universidad Nacional Aut\'{o}noma de
M\'{e}xico, 04510 M\'{e}xico, DF, MEXICO\\
$^{d}$Instituto de Investigaciones en Materiales, Universidad
Nacional Aut\'{o}noma de M\'{e}xico, 04510 M\'{e}xico, DF, MEXICO\\
$^{e}$Theoretical Division, Los Alamos National Laboratory, Los
Alamos, NM 87545, USA}

\affiliation{$^{a}$Escuela de F\'{\i}sica, Universidad Industrial
de Santander,\\
Ciudad Universitaria, Bucaramanga 680002, Colombia.\\
$^{b}$ Istituto Nazionale di Alta Matematica Francesco Severi,\\
Gruppo Nazionale di Fisica Matematica,\\
Citt$\grave{a}$ Universitaria, c.a.p. 00185, Rome, Italy.\\
$^{c}$Facultad de Ciencias, Universidad Nacional Aut\'{o}noma de
M\'{e}xico
\\
04510 M\'{e}xico, DF, MEXICO\\
$^{d}$Instituto de Investigaciones en Materiales, Universidad Nacional \\
Aut\'{o}noma de M\'{e}xico, 04510 M\'{e}xico, DF, MEXICO\\
$^{e}$Theoretical Division, Los Alamos National Laboratory, Los
Alamos, NM 87545, USA}

\address{$^{a}$Istituto Nazionale di Alta Matematica Francesco Severi \\
Gruppo Nazionale di Fisica Matematica, Citt\`{a} Universitaria, c.a.p.
00185, Rome, Italy.\\
$^{b}$Facultad de Ciencias, Universidad Nacional Aut\'{o}noma de M\'{e}xico
\\
04510 M\'{e}xico, DF, Mexico\\
$^{c}$Instituto de Investigaciones en Materiales, Universidad Nacional \\
Aut\'{o}noma de M\'{e}xico, 04510 M\'{e}xico, DF, Mexico\\
$^{c}$Theoretical Division, Los Alamos National Laboratory, Los Alamos, NM
87545, USA}

\section{Introduction}

In early works \cite{Landsberg65,Beckmann79,Beckmann82}\ on the relativistic
ideal boson gas (RIBG) explicit Bose-Einstein condensation (BEC)\ critical
transition temperature $T_{c}$-formulae were derived for both the
nonrelativistic and ultrarelativistic limits and specific-heat anomalies at $%
T_{c}$ were studied. In addition, Refs.\cite{Beckmann79,Beckmann82}
considered all space dimensions $d>0$ and delved into the relation between $%
d $ and various critical exponents. At sufficiently high temperatures,
however, boson-antiboson pair production becomes appreciable and this was
\textit{not} accounted for. The first reports to include \textit{both}
bosons and antibosons appear to be Refs.\cite{HaberWeldon81,HaberWeldon82}
where high-temperature expansions for the various thermodynamic functions
(pressure, particle-number density, entropy, specific heats, etc.) were
derived. Extensive numerical work in $d$ dimensions that does not rely on
such high-temperature expansions was reported in Refs.\cite%
{SinghPandita83,SinghPathria84}.\ In the elegant treatment of Ref.\cite%
{Goulart89}\ with inverse Mellin transforms the specific heat anomaly of the
RIBG at its BEC $T_{c}$ was found to be washed out when pair-production was
included. The relationship between the BEC of the RIBG and
spontaneous-symmetry breaking was explored in Refs.\cite%
{HaberWeldon82,Kapusta81}; see also the rather complete Ref.\cite%
{KapustaGale}, esp. \S 2.4.

BECs are also of interest in cosmological and astrophysical contexts. In
fact, increasing attention has recently been paid cosmological models that
describe dark matter (DM) as a condensate phase of some scalar boson field
\cite{cosmologicalcondensate,c1,c2,c3,c4,c5,c6,c7,c8,c9,c9 bis,c9
tris,c10,c11}. Such models are competitive with the $\Lambda $ cold dark
matter ($\Lambda $CDM) model \cite{lcdm} to explain observational properties
of DM at cosmological and astrophysical levels. In particular, a scalar
boson field with an extremely small mass of about $10^{-22}\,eV$ can explain
the cosmological evolution of the universe \cite%
{cosmologicalcondensate,c1,c2,c3,c4,c5,c6,c7,c8,c9,c9 bis,c9 tris,c10,c11},
the rotation curves of galaxies \cite{sfdm2}, the central-density profile of
low-surface-brightness galaxies \cite{sfdm3}, the size of galactic halos
\cite{sfdm4}, and the amount of substructures in the universe \cite{sfdm5}.
In Ref.\cite{briscese} it was shown that a complex and self-interacting
scalar boson field with a more realistic mass of about $1\,eV$ in a BEC is
also a viable DM candidate. Moreover, in this model no fine tuning of the
scalar-field energy density at early times is required and the condensate
formation is due to self-interactions.

Indeed, BECs are of interest in the context of quantum gravity. In Refs.\cite%
{briscese greather de llano,briscese2} some of us have shown that
Planck-scale deformations of the energy-momentum relation that naturally
emerges in many quantum-gravity theories (for an excellent nontechnical
overview see Ref.\cite{smolin}) may affect the properties of low-temperature
BECs. In particular it was shown that a Planck-scale induced deformation of
the Minkowski energy-momentum dispersion relation $E\simeq \sqrt{%
m^{2}c^{4}+p^{2}c^{2}}+\,\xi \,m\,c\,p/2M_{p}$, where $m$ is the mass of the
bosons, $M_{p}$ the Planck mass and $\xi $ a dimensionless parameter,
produces a shift in the condensation temperature $T_{c}$ of about $\Delta
T_{c}/T_{c}^{0}\simeq 10^{-6}\xi _{1}$ in typical BECs such as $_{37}^{87}$%
Rb \cite{Ander},$\ _{3}^{7}$Li \cite{Bradley},$\ _{11}^{23}$Na \cite{Davis},$%
\ _{1}^{1}$H \cite{Fried}, $_{37}^{85}$Rb \cite{Cornish}, $_{2}^{4}$He \cite%
{Pereira}, $_{19}^{41}$K \cite{Mondugno}, $_{55}^{133}$Cs \cite{Grimm}, and $%
_{24}^{52}$Cr \cite{Griesmaier}. The quantum gravity induced shift in $T_{c}$
makes possible to upper-bound the deformation parameter as $|\xi |\lesssim
10^{4}$ with recent ultra-precise measurements of $T_{c}$ as, e.g., in $%
_{19}^{39}K$ \cite{condensatePRL}. In Ref.\cite{briscese greather de
llano,briscese2} it is also discussed how to enlarge $\Delta T_{c}/T_{c}^{0}$
thus improving the bound on $\xi $ and hence realize an \textit{ad hoc}
experiment accomplish this. Finally, the Planck-scale induced shift in $T_{c}
$ is compared with similar effects due to interboson interactions and
finite-size effects. These results open a new possibility for a quantum
gravity phenomenology based on low-temperature condensates, so that BECs
truly appear to be a frontier interdisciplinary research field open to many
applications.

We also stress how the effect of interactions as well as of
finite-size effects might have observable effects on laboratory BE
condensates. For example, in Ref.\cite{condensatePRL} the effect
of interactions has been observed in $_{19}^{39}K$ and a shift in
$T_{c}$ measured as a function of the interboson s-wave scattering
length $a$ and data have been fitted with the second-order
polynomial $\Delta
T_{c}/T_{c}^{0}\simeq b_{1}(a/\lambda _{T})+b_{2}(a/\lambda _{T})^{2}$ with $%
b_{1}=-3.5\pm 0.3$ and $b_{2}=46\pm 5$, the second term being due to
beyond-mean-field effects. However, in what follows we do not consider
interaction nor finite-size effects as we focus on the ideal Bose gas. Such
effects are being investigated.

Here we study the metastability of a BEC that does not contain antibosons. A
motivation is given in \S II. In Ref.\cite{de llano} the properties a RIBG
in terms of the Helmholtz free energy with antibosons included was discussed
and shown to be a state with a lower Helmholtz free energy than that without
antibosons. In \S III we generalize this result by comparing two different
BECs, with and without antibosons, but with the same total number of
particles and at the same finite temperature. Such states are related by a
thermodynamic transformation ensuring that they are meaningfully comparable.
In particular, we rely on the law of nondecreasing entropy for isolated
systems. In \S IV we conclude that the state with antibosons has greater
entropy and lower Helmholtz free energy and which is therefore the stable
state, while the state without antibosons is metastable. In \S V we derive
the expression of the pressure of the BEC in equilibrium with a thermalized
gas of bosons and show that the state with antibosons has higher pressure.
As an overall check we calculate the Gibbs potential in both cases and show
that it is the same, as expected. Lastly, in \S VI we generalize results for
arbitrary $d>0$ dimensions, integer or not. We conclude in \S VII.

\section{Motivation}

\label{toughtexperiment1}

To study the relative stability of the two states with and without
antibosons one should compare their entropies. For a meaningful comparison
the two states must be at the same temperature, volume and number density.
This is guaranteed in what follows. Consider a system composed of two heat
reservoirs $R_{1}$ and $R_{2}$ at temperatures $T_{1}$ and $T_{2}$,
respectively, with $T_{1}\ll T_{2}$, and a gas of $N$\ bosons $B$ of mass
rest mass $m$\ contained in a volume $V$ with a number-density $n\equiv N/V$%
. The reservoirs are much larger in volume than the boson volume $V$ so they
can be placed in thermal contact with the boson gas without appreciably
changing its temperature, or they can be isolated from the boson gas. Assume
that $T_{1}<T_{c}^{B}<T_{c}^{B\bar{B}}$, where $T_{c}^{B}$ and $T_{c}^{B\bar{%
B}}$ are the boson gas BEC critical temperatures without and with antibosons
$\bar{B}$, respectively, so that at $T_{1}$ the boson gas itself is a BEC.
Assume also that $k_B T_{2}\gg mc^{2}$, with $c$ the velocity of light, and
that $k_B T_{1}\ll mc^{2}$. Initially, the gas contains only bosons and is
in thermal equilibrium with the first reservoir $R_{1}$ at temperature $%
T_{1} $. Since $k_B T_{1}\ll mc^{2}$ at this temperature any antibosons
present are negligible. The boson gas is then isolated from the reservoir $%
R_{1}$ and placed in thermal contact with the reservoir $R_{2}$. After
awhile the boson gas reaches thermal equilibrium at temperature $T_{2}$.
Since $k_B T_{2}\gg mc^{2}$ antibosons are created substantially by
pair-production so that the equilibrium state now also contains antibosons $%
\bar{B}$. Finally, the boson gas is isolated from the reservoir $R_{2}$ and
placed in thermal equilibrium with the reservoir $R_{1}$ so that the final
temperature of the boson gas is $T_{1}$. The question concerning the
metastability of the state without antibosons can be formulated in the
following way: at the end of the process just described does the boson gas
contain antibosons or does it go back to the initial state without
antibosons? To answer this one must calculate the entropy variation $\Delta
S_{tot}^{I}$ of the whole system (boson gas and reservoirs) a final state of
the boson gas with antibosons, and compare it with the entropy $\Delta
S_{tot}^{II}$ of a final state without antibosons. This question is
addressed and resolved in \S IV where we show that $\Delta
S_{tot}^{I}>\Delta S_{tot}^{II}$ so that the state without antibosons is
metastable.

We first calculate the main thermodynamic functions in both cases, with and
without antibosons.

\section{Energy density and Helmholtz free energy below BEC $T_{c}$}

We consider two gas systems, one with only bosons $B$ and a second one
containing also antibosons $\bar{B}$. They are both as the same temperature $%
T<T_{c}^{B}<T_{c}^{B\bar{B}}$, where $T_{c}^{B}$ and $T_{c}^{B\bar{B}}$ are
the condensation temperatures of these two systems without and with
antibosons, respectively. We first write down explicit expressions for
internal energies and number densities and then proceed to calculate their
Helmholtz free energy.\textbf{\ }

Since $T<T_{c}^{B}<T_{c}^{B\bar{B}}$ the condensate forms in both a system
containing only bosons $B$ as well as in a system containing also antibosons$%
\ \bar{B}$. At such temperatures $T$ the chemical potential $\mu \simeq
mc^{2}$ in a RIBG whose energy $E(p)$-momentum $p$ dispersion is $E(p)\equiv
\sqrt{p^{2}c^{2}+m^{2}c^{4}}$. For a gas containing only bosons the number
density is%
\begin{equation}
n=n_{0}+(\hbar ^{3}2\pi ^{2})^{-1}\int_{0^{+}}^{\infty }p^{2}dp\frac{1}{\exp
{\left[ \beta (E(p)-mc^{2})\right] }-1}\,  \label{numberdensityB}
\end{equation}
where $\beta \equiv 1/k_{B}T.$ The net internal energy per unit volume $V$\
is

\begin{equation}
\frac{U^{B}(n,T,V)}{V}=mc^{2}n_{0} +(\hbar ^{3}2\pi
^{2})^{-1}\int_{0^{+}}^{\infty }p^{2}dp\frac{E(p)}{\exp {\ \left[ \beta
(E(p)-mc^{2})\right] }-1}.
\end{equation}
Here
\begin{equation}
n_{0}\equiv \frac{1}{V}\frac{1}{\exp [\beta (mc^{2}-\mu )]-1}.  \label{n0}
\end{equation}
Combining these equations leaves

\begin{equation}
\frac{U^{B}(n,T,V)}{V}=mc^{2}n +(\hbar ^{3}2\pi
^{2})^{-1}\int_{0^{+}}^{\infty }p^{2}dp\frac{E(p)-mc^{2}}{\exp {\left[ \beta
(E(p)-mc^{2})\right] }-1}.  \label{energycondensateB}
\end{equation}

When antibosons are included the number density $n$ is

\begin{equation}
n=n_{0}+(\hbar ^{3}2\pi ^{2})^{-1}\int_{0^{+}}^{\infty }p^{2}dp
\left[ \frac{1}{\exp {\left[ \beta (E(p)-mc^{2})\right]
}-1}-\frac{1}{ \exp {\left[ \beta (E(p)+mc^{2})\right] }-1}\right]
\label{numberdensityBB}
\end{equation}
so that

\begin{equation}
\frac{U^{B\bar{B}}(n,T,V)}{V}=mc^{2}n_{0}+(\hbar ^{3}2\pi
^{2})^{-1}\int_{0^{+}}^{\infty }p^{2}dp\,E(p) \left[ \frac{1}{\exp {%
\left[ \beta (E(p)-mc^{2})\right] }-1}+\frac{1}{ \exp {\left[ \beta
(E(p)+mc^{2})\right] }-1}\right] .  \label{numberdensityBB}
\end{equation}
Combining these two equations gives

\begin{equation}
\frac{U^{B\bar{B}}(n,T,V)}{V}=mc^{2}n+(\hbar ^{3}2\pi
^{2})^{-1}\int_{0^{+}}^{\infty }p^{2}dp \left[ \frac{E(p)-m\,c^{2}}{%
\exp {\left[ \beta (E(p)-mc^{2})\right] } -1}+\frac{E(p)+m\,c^{2}}{\exp {%
\left[ \beta (E(p)+mc^{2})\right] }-1}\right] .  \label{energycondensateBB}
\end{equation}

The Helmholtz free energy per unit volume without antibosons is then

\begin{equation}
F^{B}(T,V,n)/V=mc^{2}n+k_{B}T(\hbar ^{3}2\pi
^{2})^{-1}\int_{0^{+}}^{\infty }p^{2}dp \ln \left[ 1-\exp \left(
\beta \left[ mc^{2}-E(p)\right] \right) \right] .
\label{HelmotzBII}
\end{equation}%
In the case with antibosons one has

\[
F^{B\bar{B}}(T,V,n)/V=mc^{2}n+k_{B}T(\hbar ^{3}2\pi
^{2})^{-1}\int_{0^{+}}^{\infty }p^{2}dp \{\ln \left[ 1-\exp \left[
\beta \left( mc^{2}-E(p)\right) \right] \right] +
\]

\begin{equation}
\ln \left[ 1-\exp \left[ -\beta \left( mc^{2}+E(p)\right) \right] \right] \}.
\label{HelmotzBBII}
\end{equation}

From (\ref{HelmotzBII})-(\ref{HelmotzBBII}) it also follows that

\begin{equation}
F^{B\bar{B}}(T,V,n)-F^{B}(T,V,n)= V \frac{k_{B}T}{\hbar ^{3}2\pi
^{2}} \int_{0^{+}}^{\infty }p^{2}dp \ln \left[ 1-\exp \left(
-\beta \left[ mc^{2}+E(p)\right] \right) \right] <0.
\label{helmotzdifference}
\end{equation}
Therefore the state containing antibosons has a lower Helmholtz free energy.
This same result was found in Ref.\cite{de llano} except that here the
Helmholtz free energies are compared at the same temperature $T$.

\section{Entropy}

\label{section entropy}

Here we calculate the entropy of the boson field with and without
antibosons. This is then used to determine the entropy variation in the
thermodynamic transformation described in \S \ref{toughtexperiment1} to
conclude that the state containing antibosons is the stable state while the
state without antibosons is only metastable. If only $B$ bosons are
considered, the entropy follows from

\begin{equation}
TS^{B}(T,V,n)=U^{B}(T,V,n)-F^{B}(T,V,n)  \label{entropytc}
\end{equation}%
where the internal energy per unit volume is given by (\ref%
{energycondensateB}). Whence

\begin{equation}
S^{B}(T,V,n)/V=k_{B}\,(\hbar ^{3}2\pi
^{2})^{-1}\int_{0^{+}}^{\infty }p^{2}dp {\Huge \{}\frac{\beta
(E(p)-mc^{2})}{\exp {\left[ \beta
(E(p)-mc^{2})\right] }-1}-\ln \left[ 1-\exp \left( \beta \left[ mc^{2}-E(p)%
\right] \right) \right] {\huge \}.}  \label{entropyBII}
\end{equation}%
If antibosons are included the entropy follows from%
\begin{equation}
TS^{B\bar{B}}(T,V,n)=U^{B\bar{B}}(T,V,n)-F^{B\bar{B}}(T,V,n)
\label{entropytc}
\end{equation}%
where the Helmholtz free energy is given by (\ref{HelmotzBBII}). Using the
latter and (\ref{energycondensateBB}) one gets

\[
S^{B\bar{B}}(T,V,n)/V=k_{B}\,(\hbar ^{3}2\pi ^{2})^{-1}\int_{0^{+}}^{\infty
}p^{2}dp \{\frac{\beta (E(p)-mc^{2})}{\exp {\left[ \beta (E(p)-mc^{2})%
\right] } -1}+\frac{\beta (E(p)+mc^{2})}{\exp {\left[ \beta (E(p)+mc^{2})%
\right] }-1}+
\]

\begin{equation}
-\ln \left[ 1-\exp \left[ \beta \left( mc^{2}-E(p)\right) \right] \right]
-\ln \left[ 1-\exp \left[ -\beta \left( mc^{2}+E(p)\right) \right] \right]
\}.  \label{entropyBBII}
\end{equation}%
One can nowcompare the entropies of the two states with and without
antibosons. From (\ref{entropyBII}) and (\ref{entropyBBII}) one easily finds
that

\[
S^{B\bar{B}}(T,V,n)-S^{B}(T,V,n)=k_{B}V(\hbar ^{3}2\pi
^{2})^{-1}\int_{0^{+}}^{\infty }p^{2}dp {\LARGE \{}\beta \left(
E(p)+mc^{2}\right) /[\exp {\left[ \beta (E(p)+mc^{2})\right] }-1]+
\]

\begin{equation}
-\ln \left[ 1-\exp \left( -\beta \left[ mc^{2}+E(p)\right] \right) \right]
{\LARGE \}}>0  \label{entropydifference}
\end{equation}%
so that the state \textit{without} antibosons being less entropic is thus
metastable.

Now consider the thermodynamic transformation described in \S \ref%
{toughtexperiment1} and calculate the total entropy variation of the boson
field plus that of the two reservoirs. This enables one to decide if the
final state will contain or not antibosons. The final state of the whole
system (boson field plus reservoirs) turns out to be more entropic one which
in turn implies that the state with antibosons is the stable state while the
state without antibosons is only metastable. If the final state of the gas
also contains antibosons, the entropy variation of the gas in the
thermodynamic transformation described in \S \ref{toughtexperiment1} is

\begin{equation}
\Delta S_{gas}^{I}=S^{B\bar{B}}(T_{1},V,n)-S^{B}(T_{1},V,n)>0
\label{entropydifference2}
\end{equation}%
(which is the same as (\ref{entropydifference})), while the
entropy variation of the two reservoirs is

\begin{equation}
\Delta S_{1}^{I}=\frac{\Delta \mathit{Q}_{1}}{T_{1}}=\frac{U^{B\bar{B}
}(T_{2},V,n)-U^{B}(T_{1},V,n)}{T_{1}}>0
\end{equation}

\begin{equation}
\Delta S_{2}^{I}=\frac{\Delta \mathit{Q}_{2}}{T_{2}}=\frac{U^{B\bar{B}%
}(T_{1},V,n)-U^{B\bar{B}}(T_{2},V,n)}{T_{2}}<0.
\end{equation}%
Hence, the total entropy variation is

\[
\Delta S_{tot}^{I}=\Delta S_{gas}^{I}+\Delta S_{1}^{I}+\Delta S_{2}^{I}
=\Delta S_{gas}^{I}+U^{B\bar{B}}(T_{2},V,n)\left( \frac{1}{T_{1}}-\frac{1}{%
T_{2}}\right) + \frac{U^{B\bar{B}}(T_{1},V,n)}{T_{2}}-\frac{U^{B}(T_{1},V,n)%
}{T_{1}}>
\]

\begin{equation}
>\Delta S_{gas}^{I}+\left[ U^{B\bar{B}}(T_{2},V,n)-U^{B}(T_{1},V,n)\right]
\left( \frac{1}{T_{1}}-\frac{1}{T_{2}}\right) >0.
\end{equation}%
The transformation is thus allowed but is irreversible.

The net entropy variation in the thermodynamic transformation of \S \ref%
{toughtexperiment1} when the final state is without antibosons is thus

\begin{equation}
\Delta S_{gas}^{II}=0
\end{equation}%
while the entropy variation of the two reservoirs is

\begin{equation}
\Delta S_{1}^{II}=\frac{U^{B\bar{B}}(T_{2},V,n)-U^{B}(T_{1},V,n)}{T_{1}}>0
\end{equation}

\begin{equation}
\Delta S_{2}^{II}=\frac{U^{B}(T_{1},V,n)-U^{B\bar{B}}(T_{2},V,n)}{T_{2}}<0.
\end{equation}
Hence, the total entropy variation is

\begin{equation}
\begin{array}{ll}
\Delta S_{tot}^{II}=\left[ U^{B\bar{B}}(T_{2},V,n)-U^{B}(T_{1},V,n)\right]
\left( \frac{1}{T_{1}}-\frac{1}{T_{2}}\right) >0. &
\end{array}
\label{deltaSIItot}
\end{equation}%
Again, the transformation is allowed but is irreversible and its only effect
is a heat transfer between the two reservoirs $R_{1}$ and $R_{2}$.

We can now compare the two entropy variations $\Delta S_{tot}^{I}$ and $%
\Delta S_{tot}^{II}$. One has

\begin{equation}
\Delta S_{tot}^{I}-\Delta S_{tot}^{II}=\Delta S_{gas} +\frac{U^{B\bar{B}%
}(T_{1},V,n)-U^{B}(T_{1},V,n)}{T_{2}}>0  \label{totalDS}
\end{equation}
so that $\Delta S_{tot}^{I}>\Delta S_{tot}^{II}$, i.e., the entropy
variation is greater in the case in which the final state contains
antibosons. Therefore, the final state of the whole system of the gas plus
the two reservoirs is more entropic when the gas contains antibosons in the
final state. Again this means the state without antibosons is metastable and
that the final equilibrium state described in \S \ref{toughtexperiment1} is
the one containing antibosons.

We remark that if the boson field were a \textit{real} scalar field it would
not admit antibosons and in this instance the state containing only bosons $%
B $ is the only possible one and therefore it is not metastable but rather a
stable state. One thus concludes that if antibosons are allowed, namely if
the scalar field is \textit{complex} as assumed here, the state with
antibosons \textit{is} allowed and this state will be the stable one while
the state without antibosons will only be metastable.

\section{Pressure}

Following the same procedure we introduce the pressure $P$ as a function of $%
n$, $V$ and $T$ . Specifically, if no antibosons are present

\begin{equation}
\beta VP^{B}=-\ln \left[ 1-\exp \left[ \beta (\mu -mc^{2})\right]
\right] -(\hbar ^{3}2\pi ^{2})^{-1}V \int_{0^{+}}^{\infty
}p^{2}dp\ln \left[ 1-\exp \left[ \beta (\mu -E(p))\right] \right]
.  \label{PV B}
\end{equation}
We rewrite (\ref{numberdensityB}) as

\begin{equation}
n_{0}=n-n_{+}=\frac{1}{V}\frac{\exp [\beta (\mu -mc^{2})]}{1-\exp [\beta
(\mu -mc^{2})]}  \label{n0B}
\end{equation}%
where

\begin{equation}
n_{+}\equiv (\hbar ^{3}2\pi ^{2})^{-1}\int_{0^{+}}^{\infty }p^{2}dp\frac{1}{%
\exp {\left[ \beta (E(p)-mc^{2})\right] }-1}
\end{equation}%
is the number density of noncondensate (or excited) bosons. Below the
condensation temperature $\mu \simeq mc^{2}$ so that $\exp \left[ \beta (\mu
-mc^{2})\right] \simeq 1$ apart from small corrections $O(1/V)$ which vanish
in the thermodynamic limit $V\rightarrow \infty $. We can write the
logarithm in the rhs of (\ref{PV B}) as

\begin{equation}
\ln \left[ 1-\exp \left[ \beta (\mu -mc^{2})\right] \right] \simeq -\ln %
\left[ V(n-n_{+})\right]
\end{equation}
whence

\[
\beta VP^{B}=-\ln \left[ V\left( n-\frac{1}{(\hbar ^{3}2\pi
^{2})}\right. \right.  \left. \left. \int_{0^{+}}^{\infty
}p^{2}dp\frac{1}{\exp \left[ \beta \left( E(p)-mc^{2}\right)
\right] -1}\right) \right]
\]

\begin{equation}
-\frac{V}{(\hbar ^{3}2\pi ^{2})}\int_{0^{+}}^{\infty }p^{2}dp\ln
\left[ 1-\exp \left[\beta \left( mc^{2}-E(p)\right) \right]\right]
.
\end{equation}%
Dividing through by $V$ gives the first term on the rhs proportional to $%
V^{-1}\ln V$ which also vanishes in the thermodynamic limit, so one gets
\begin{equation}
P^{B}=-\frac{k_{B}T}{(\hbar ^{3}2\pi ^{2})}\int_{0^{+}}^{\infty }p^{2}dp\ln %
\left[ 1-\exp [\beta \left( mc^{2}-E(p)\right) ]\right] .  \label{PV B final}
\end{equation}

When antibosons are included the pressure is given by the relation

\[
\beta VP^{B\bar{B}}=-\ln \left[V n-\frac{1}{\hbar ^{3}2\pi ^{2}}
\int_{0^{+}}^{\infty }p^{2}dp \left( \frac{1}{\exp \left[ \beta
(E(p)-mc^{2})\right] -1}-\frac{1}{\exp \left[ \beta (E(p)+mc^{2})\right] -1}%
\right) \right]+
\]

\begin{equation}
-\frac{V}{(\hbar ^{3}2\pi ^{2})}\int_{0^{+}}^{\infty }p^{2}dp\left[ \ln
\left( 1-\exp [\beta \left( mc^{2}-E(p)\right) ]\right) +\ln \left( 1-\exp
[-\beta \left( mc^{2}+E(p)\right) ]\right) \, \, \right] .  \label{PV BB}
\end{equation}%
Since the first term in (\ref{PV BB}) is negligible in the thermodynamic
limit, the final result is

\begin{equation}
P^{B\bar{B}}=-\frac{k_{B}T}{(\hbar ^{3}2\pi ^{2})}\int_{0^{+}}^{\infty
}p^{2}dp\left( \ln \left[ 1-\exp \left[ \beta \left( mc^{2}-E(p)\right) %
\right] \right] +\ln \left[ 1-\exp \left[ -\beta \left( mc^{2}+E(p)\right) %
\right] \right] \right) .  \label{PV BB final}
\end{equation}%
Comparing (\ref{PV B final}) with (\ref{PV BB final}) it becomes evident
that when antibosons are included the pressure is greater than in the case
without antibosons. Indeed, one has

\begin{equation}
P^{B\bar{B}}-P^{B}=-\frac{k_{B}T}{(\hbar ^{3}2\pi ^{2})}\int_{0^{+}}^{\infty
}p^{2}dp \ln \left[ 1-\exp \left[ -\beta \left( mc^{2}+E(p)\right) %
\right] \right] >0.  \label{Delta
PV}
\end{equation}

As a final overall check, comparison of equations (\ref{helmotzdifference})
and (\ref{Delta PV}) shows that the state with and without antibosons have
the same Gibbs free energy $G(P,T)=F+PV=\mu N$, namely $G^{B\bar{B}%
}(P,T)=G^{B}(P,T)$, as must be the case since the net number of particles $N$
and chemical potential $\mu $\ are the same.

\section{Generalization to $d$ spatial dimensions}

Here we generalize the thermodynamic potentials to arbitrary $d>0$ spatial
dimensions, integer or not. The result is confirmed that the state with only
bosons is metastable while the state with both bosons and antibosons is
stable. To motivate this section we recall that spaces with dimensionality
different from $d=3$ are considered in many physical contexts, e.g., in
quantum gravity (see Ref.\cite{smolin} for a review). In other areas, e.g.,
Mandelbrot (Ref.\cite{mandelbrot}, p. 85) cites an empirical fractal
dimension $d=1.23$ for the distribution of galaxies in the observable
universe.

We first calculate the thermodynamic functions with and without
antibosons. Assuming a real nonnegative number $d$ of spatial
dimension, the
sum over momentum now becomes%
\begin{equation}
\sum_{\mathbf{p}\neq 0}\longrightarrow \left( \frac{L}{2\pi \hbar }\right)
^{d}\,\Omega _{d}\int_{0^{+}}^{\infty }p^{d-1}\,dp
\end{equation}%
where $\Omega _{d}$ is the solid angle in $d$ dimensions and the system
volume is $L^{d}$.

If no antibosons are present the number density $n$ is%
\begin{equation}
n=n_{0}+\Omega _{d}\,(2\pi \hbar )^{-d}\,\int_{0^{+}}^{\infty }p^{d-1}dp\,
\left[ \exp \left[ \beta \left( E(p)-\mu \right) \right] -1\right] ^{-1}
\end{equation}%
where%
\begin{equation}
n_{0}\equiv \left[ V\left( \exp \left[ \beta \left( mc^{2}-\mu \right) %
\right] -1\right) \right] ^{-1}.  \label{numberdensitymodified}
\end{equation}%
As before, $n_{0}\equiv N_{0}/L^{d}$ is the number density of zero-momentum $%
\mathbf{p}=0$\ bosons within a $d$-dimensional volume $V\equiv L^{d}$. The
internal energy per unit volume is%
\begin{equation}
U^{B}(T,V,n)/V=nmc^{2}+\Omega _{d}\left( 2\pi \hbar \right) ^{-d}
\int_{0^{+}}^{\infty }p^{d-1}dp\,\frac{E(p)-mc^{2}}{\exp \left[
\beta \left( E(p)-\mu \right) \right] -1}.  \label{d1}
\end{equation}%
The Helmholtz free energy is%
\begin{equation}
F^{B}(T,V,n)/V=mc^{2}n+k_{B}T\,\Omega _{d}\left( 2\pi \hbar
\right) ^{-d} \int_{0^{+}}^{\infty }p^{d-1}dp\,\ln \left[ 1-\exp
\left[ \beta \left( \mu -E(p)\right) \right] \right]   \label{d2}
\end{equation}%
while the entropy per unit volume is now%
\begin{equation}
S^{B}(T,V,n)/V=k_{B}\Omega _{d}\,\left( 2\pi \hbar \right)
^{-d}\int_{0^{+}}^{\infty }\,p^{d-1}dp\{\frac{\beta \left(
E(p)-mc^{2}\right) }{\exp \left[ \beta \left( E(p)-\mu \right) \right] -1}%
-\ln \left[ 1-\exp \left[ \beta \left( \mu -E(p)\right) \right] \right] \}
\label{d3}
\end{equation}%
with the term $\left( k_{B}/V\right) \ln \left[ 1-\exp \left[ \beta \left(
\mu -mc^{2}\right) \right] \right] $ being negligible in the thermodynamic
limit.

If antibosons are present the number density is%
\begin{equation}
n=n_{0}+\Omega _{d}\,(2\pi \hbar )^{-d}\int_{0^{+}}^{\infty
}\,p^{d-1}\,dp \left[ \frac{1}{\exp \left[ \beta (E(p)-\mu )\right] -1}%
+\frac{1}{\exp \left[ \beta (E(p)+\mu )\right] -1}\right]
\end{equation}%
where $n_{0}$ is still (\ref{numberdensitymodified}). The internal energy
per unit volume is%
\begin{equation}
\frac{U^{B\bar{B}}(n,T,V)}{V}=mc^{2}n+\Omega _{d}\,\left( 2\pi \hbar \right)
^{-d}\int_{0^{+}}^{\infty }p^{d-1}dp \left[ \frac{E(p)-m\,c^{2}}{\exp {%
\left[ \beta (E(p)-mc^{2})\right] }-1}+\frac{E(p)+m\,c^{2}}{\exp {\left[
\beta (E(p)+mc^{2})\right] }-1}\right] .  \label{d4}
\end{equation}%
The Helmholtz free energy per unit volume is%
\[
F^{B\bar{B}}(T,V,n)/V=mc^{2}n+k_{B}T\Omega _{d}\,\left( 2\pi \hbar
\right) ^{-d}\int_{0^{+}}^{\infty }p^{d-1}dp \{\ln \left[ 1-\exp
\left[ \beta \left( mc^{2}-E(p)\right) \right] \right] +
\]

\begin{equation}
\ln \left[ 1-\exp \left[ -\beta \left( mc^{2}+E(p)\right) \right] \right] \}.
\label{d5}
\end{equation}

Finally, the entropy per unit volume becomes

\[
S^{B\bar{B}}(T,V,n)/V=k_{B}\,\Omega _{d}\,\left( 2\pi \hbar \right)
^{-d}\int_{0}^{\infty }p^{d-1}dp \{\frac{\beta (E(p)-mc^{2})}{\exp {%
\left[ \beta (E(p)-mc^{2})\right] } -1}+\frac{\beta (E(p)+mc^{2})}{\exp {%
\left[ \beta (E(p)+mc^{2})\right] }-1}+
\]

\begin{equation}
-\ln \left[ 1-\exp \left[ \beta \left( mc^{2}-E(p)\right) \right] \right]
-\ln \left[ 1-\exp \left[ -\beta \left( mc^{2}+E(p)\right) \right] \right]
\}.  \label{d6}
\end{equation}

At this point it is easy to generalize to arbitrary $d$ the result
that the state with only bosons is stable, just generalizing
(\ref{totalDS}). For example, by use of
(\ref{d1}-\ref{d2}-\ref{d3}-\ref{d4}-\ref{d5}-\ref{d6}) one easily
generalizes (\ref{helmotzdifference}) and obtains
\begin{equation}
F^{B\bar{B}}(T,V,n)-F^{B}(T,V,n)=\frac{V_{d}\Omega _{d}\,k_{B}T}{(2\pi \hbar
)^{d}}\int_{0^{+}}^{\infty }p^{d-1}dp \ln \left[ 1-\exp \left( -\beta %
\left[ mc^{2}+E(p)\right] \right) \right] <0
\end{equation}%
for the difference of the Helmholtz potential. Again, for
arbitrary $d$ the state containing antibosons has a lower
Helmholtz free energy. Then one can generalize
(\ref{entropydifference}) and obtain\textbf{\
\[
S^{B\bar{B}}(T,V,n)-S^{B}(T,V,n)=\frac{V_{d}\Omega
_{d}\,k_{B}}{(2\pi \hbar )^{d}}\int_{0^{+}}^{\infty }p^{d-1}dp
{\LARGE \{}\beta \left( E(p)+mc^{2}\right) /[\exp {\left[ \beta
(E(p)+mc^{2})\right] }-1]+
\]%
}

\textbf{%
\begin{equation}
-\ln \left[ 1-\exp \left( -\beta \left[ mc^{2}+E(p)\right] \right) \right]
{\LARGE \}}>0  \label{d-entropydifference}
\end{equation}%
}so that the state with antibosons is more entropic also for arbitrary $d$.

Proceeding in the same way one verifies that the relations resumed in Eq.s(%
\ref{entropydifference}-\ref{deltaSIItot}) are still valid and therefore Eq.(%
\ref{totalDS}) is also valid for arbitrary $d$. Therefore one concludes
that, also for arbitrary $d$, the state without antibosons is metastable.

\section{Conclusions}

The metastability of a Bose-Einstein condensate (BEC) that does not contain
antibosons was studied for the relativistic ideal Bose gas (RIBG). In
particular, the Helmholtz free energy with both bosons and antibosons was
shown to be a state with a lower Helmholtz potential than that without
antibosons. This was done with the same number of particles and at the same
finite temperature below the BEC critical temperature. Both states were
found to be related by a thermodynamic transformation ensuring that they are
meaningfully comparable. In addition, relying on the principle of
nondecreasing entropy for isolated systems we found that the state with
antibosons has greater entropy and is therefore the stable state, while the
state without antibosons is metastable. The pressure of both systems was
calculated and found to be higher for the state with antibosons than for the
state without them. We also confirm that the two states with and without
antibosons have the same Gibbs free energy, as expected. Lastly, results
were generalized for arbitrary dimensions $d>0$, integer or not.

\textbf{Acknowledgements}: This work was completed during a visit of FB at
UNAM-IIM in Mexico City. MdeLl thanks UNAM-DGAPA-PAPIIT (M\'{e}xico) for
grant IN102011. This work was supported in part by the U.S. Energy
Department at the Los Alamos National Laboratory. F. Briscese is a Marie
Curie fellow of the Istituto Nazionale di Alta Matematica Francesco Severi.


\begin{thebibliography}{99}
\bibitem{Landsberg65} P.T. Landsberg and J. Dunning-Davies, Phys. Rev. A
\textbf{138}, 1049 (1965).

\bibitem{Beckmann79} R. Beckmann, F. Karsch, and D.E. Miller, Phys. Rev.
Lett. \textbf{43}, 1277 (1979).

\bibitem{Beckmann82} R. Beckmann, F. Karsch, and D.E. Miller, Phys. Rev. A
\textbf{25}, 561 (1982).

\bibitem{HaberWeldon81} H.E. Haber and H.A. Weldon, Phys. Rev. Lett. \textbf{%
46}, 1497 (1981).

\bibitem{HaberWeldon82} H.E. Haber and H.A. Weldon, Phys. Rev. D \textbf{25}%
, 502 (1982).

\bibitem{SinghPandita83} S. Singh and P.N. Pandita, Phys. Rev. A \textbf{28}%
, 1752 (1983).

\bibitem{SinghPathria84} S. Singh and R.K. Pathria, Phys. Rev. A \textbf{30}%
, 442 (1984); Phys. Rev. A \textbf{30}, 3198 (1984).

\bibitem{Goulart89} H.O. Frota, M.S. Silva, and S. Goulart Rosa, Jr., Phys.
Rev. A \textbf{39}, 830 (1989).

\bibitem{Kapusta81} J.I. Kapusta, Phys. Rev. D \textbf{24}, 426 (1981).

\bibitem{KapustaGale} J.I. Kapusta and C. Gale, \textit{Finite Temperature
Field Theory: Theory and Applications}, 2nd Ed. (Cambridge University Press,
Cambridge, UK, 2006).

\bibitem{cosmologicalcondensate} A.P. Lundgren, M. Bondarescu, R.
Bondarescu, and J. Balakrishna, Astrophys. J. \textbf{715}, L35 (2010).

\bibitem{c1} I. Rodriguez-Montoya, J. Maga\~{n}a, T. Matos, and A.P\'{e}%
rez-Lorenzana, Astrophys. J. \textbf{721}, 1509 (2010).

\bibitem{c2} T.P. Woo and T. Chiueh, Astrophys. J. \textbf{697}, 850 (2009).

\bibitem{c3} L.A. Ure\~{n}a-L\'{o}pez, JCAP \textbf{0901}, 014 (2009).

\bibitem{c4} S. Fagnocchi, S. Finazzi, S. Liberati, M. Kormos, and A.
Trombettoni, New J. Phys. \textbf{12}, 095012 (2010).

\bibitem{c5} T. Harko, Monthly Not. Roy. Astron. Soc. \textbf{413}, 3095
(2011).

\bibitem{c6} A. Su\'{a}rez and T. Matos, arXiv:1101.4039 [gr-qc].

\bibitem{c7} T. Harko and F.S.N. Lobo, arXiv:1104.2674 [gr-qc].

\bibitem{c8} L.A. Gergely, T. Harko, M. Dwornik, G. Kupi, and Z. Keresztes,
arXiv:1105.0159 [gr-qc].

\bibitem{c9} T. Harko, Phys. Rev. D \textbf{83}, 123515 (2011).

\bibitem{c9 bis} T.~Harko, Mon.\ Not.\ Roy.\ Astron.\ Soc.\ \textbf{413}
(2011) 3095.

\bibitem{c9 tris} P.H. Chavanis, Phys.Rev. D \textbf{84} 043531 (2011).

\bibitem{c10} J. Barranco, A. Bernal, J.C. Degollado, A. Diez-Tejedor, M.
Megevand, M. Alcubierre, D. N\'{u}\~{n}ez, and O. Sarbach, arXiv:1108.0931
[gr-qc].

\bibitem{c11} J. Barranco and A. Bernal, arXiv:1108.1208 [astro-ph.CO].

\bibitem{lcdm} E. Komatsu \textit{et al., Seven-Year Wilkinson Microwave
Anisotropy Probe (WMAP) Observations: Cosmological Interpretation.}
arXiv:1001.4538v2 [astro-ph.CO].

\bibitem{sfdm2} C.G. Boehmer and T. Harko, JCAP \textbf{0706}, 025 (2007).

\bibitem{sfdm3} A. Bernal, T. Matos, and D. N\'{u}\~{n}ez, Rev. Mex. A. A.
\textbf{44}, 149 (2008). arXiv:astro-ph/0303455.


\bibitem{sfdm4} M. Alcubierre, F. S. Guzman, T. Matos, D. N\'{u}\~{n}ez, L.
A. Ure\~{n}a, and P. Wiederhold, Class. Quant. Grav. \textbf{19}, 5017
(2002).

\bibitem{sfdm5} T. Matos and L. A. Ure\~{n}a. Phys Rev. D \textbf{63},
063506 (2001).

\bibitem{briscese} F. Briscese, Phys. Lett. B, \textbf{696}, 315 (2011).

\bibitem{Ander} M.H. Anderson, J.R. Ensher, M.R. Wieman, and E.A. Cornell,
Science \textbf{269}, 198 (1995).

\bibitem{Bradley} C.C. Bradley, C.A. Sackett, J.J. Tollett, and R.G. Hulet,
Phys. Rev. Lett. \textbf{75}, 1687 (1995).

\bibitem{Davis} K.B. Davis, M.O. Mewes, M.R. Andrews, N.J. van Drutten, D.S.
Durfee, D.M. Kurn, and W. Ketterle, Phys. Rev. Lett. \textbf{75}, 3969
(1995).

\bibitem{Fried} D.G. Fried, T.C. Killian, L. Willmann, D. Landhuis, S.C.
Moss, D. Kleppner, and T.J. Greytak, Phys. Rev. Lett. \textbf{81}, 3811
(1998). 

\bibitem{Cornish} S.L. Cornish, N.R. Claussen, J.L. Roberts, E.A. Cornell,
and C.E. Wieman, Phys. Rev. Lett. \textbf{85}, 1795 (2000). 

\bibitem{Pereira} F. Pereira Dos Santos, J. L\'{e}onard, Junmin Wang, C.J.
Barrelet, F. Perales, E. Rasel, C.S. Unnikrishnan, M. Leduc, and C.
Cohen-Tannoudji, Phys. Rev. Lett. \textbf{86}, 3459 (2001). 

\bibitem{Mondugno} G. Mondugno, G. Ferrari, G. Roati, R.J. Brecha, A.
Simoni, and M. Inguscio, Science \textbf{294}, 1320 (2001).

\bibitem{Grimm} T. Weber, J. Herbig, M. Mark, H.C. Nagel, and R. Grimm,
Science \textbf{299}, 232 (2003).

\bibitem{Griesmaier} A. Griesmaier, J. Werner, S. Hensler, J. Stuhler, and
T. Pfau, Phys. Rev. Lett. \textbf{94}, 160401 (2005).

\bibitem{condensatePRL} R.P. Smith, R.L.D. Campbell, N. Tammuz, and Z.
Hadzibabic, Phys. Rev. Lett. \textbf{106}, 250403 (2011).

\bibitem{de llano} M. Grether, M. de Llano, and G.A. Baker, Jr., Phys. Rev.
Lett. \textbf{99}, 200406 (2007).

\bibitem{briscese greather de llano} F. Briscese, M. Grether, and M. de
Llano, Europhys. Lett. \textbf{98} (2012) 60001.

\bibitem{briscese2} F. Briscese, arXiv:1206.1236 [gr-qc].

\bibitem{smolin} L. Smolin, \textit{Three Roads to Quantum Gravity} (Basic
Books, NY, 2002).

\bibitem{mandelbrot} B.B. Mandelbrot, \textit{The Fractal Geometry of Nature}
(W.H. Freeman, San Francisco, 1982).
\end{thebibliography}
\end{document}